\documentclass[12pt]{article} 
\pdfoutput=1

\usepackage[utf8]{inputenc}
\usepackage{jheppub}
\usepackage{amsmath}
\usepackage{amsfonts}
\usepackage{amssymb}
\usepackage{mathtools}
\usepackage{amsthm}
\usepackage{dsfont}
\usepackage{hyperref}
\usepackage[capitalise,sort]{cleveref}
\usepackage{pgfplots}
\usepackage{tabularx}
\usepackage{environ}
\usepackage[inline]{enumitem}
\usepackage{comment}
\usepackage{tikz}
\usetikzlibrary{decorations.pathmorphing,calc,fit,matrix,cd,shapes,arrows,positioning,chains,positioning,arrows.meta,hobby,patterns}
\usepackage[
style=numeric-comp
,sorting=none
,maxbibnames=99
,giveninits=true
,backend=bibtex 
]{biblatex}
\usepackage{blkarray}
\usepackage{shadow}
\usepackage{fancybox}
\usepackage{bm}
\usepackage{abstract}
\usepackage{lscape}
\usepackage{verbatim}
\usepackage[normalem]{ulem}
\usepackage{subcaption}
\DeclareFieldFormat{doilink}{
\iffieldundef{doi}{#1}
{\href{http://dx.doi.org/\thefield{doi}}{#1}}}

\DeclareBibliographyDriver{article}{%
  \usebibmacro{bibindex}%
  \usebibmacro{begentry}%
  \usebibmacro{author/translator+others}%
  \setunit{\labelnamepunct}\newblock
  \usebibmacro{title}%
  \newunit
  \printlist{language}%
  \newunit\newblock
  \usebibmacro{byauthor}%
  \newunit\newblock
  \usebibmacro{bytranslator+others}%
  \newunit\newblock
  \printfield{version}%
  \newunit\newblock
  \usebibmacro{in:}%
  \href{http://dx.doi.org/\thefield{doi}}{%
  \usebibmacro{journal+issuetitle}%
  \newunit\newblock
  \usebibmacro{byeditor+others}%
  \newunit\newblock
  \usebibmacro{note+pages}}%
  \newunit\newblock
  \iftoggle{bbx:isbn}
   {\printfield{issn}}
   {}%
  \newunit\newblock
  \usebibmacro{doi+eprint+url}%
  \newunit\newblock
  \usebibmacro{addendum+pubstate}%
  \newunit\newblock
  \usebibmacro{pageref}%
  \usebibmacro{finentry}
  }
  
\DeclareFieldFormat{url}{\mkbibacro{URL}\addcolon\space\href{#1}{Link}}

\pgfplotsset{
        compat=1.9,
        compat/bar nodes=1.8,
    }

\makeatletter
\def\@xfootnote[#1]{%
	\protected@xdef\@thefnmark{#1}%
	\@footnotemark\@footnotetext}
\makeatother

\definecolor{prhigh}{HTML}{ff0000}
\definecolor{sechigh}{HTML}{e0fbfc}
\definecolor{prcolor}{HTML}{1d3557}
\definecolor{seccolor}{HTML}{457b9d}
\definecolor{tercolor}{HTML}{98c1d9}

\newcommand\bea{\begin{eqnarray}}
\newcommand\eea{\end{eqnarray}}

\theoremstyle{plain}

\theoremstyle{definition}

\theoremstyle{remark}



\renewcommand{\epsilon}{\varepsilon}

\makeatletter
\newsavebox{\measure@tikzpicture}
\NewEnviron{scaletikzpicturetowidth}[1]{%
  \def\tikz@width{#1}%
  \begin{lrbox}{\measure@tikzpicture}%
  \BODY
  \end{lrbox}%
  \pgfmathparse{#1/\wd\measure@tikzpicture}%
  \BODY
}
\makeatother

\makeatletter
\newcommand{\inlineitem}[1][]{%
\ifnum\enit@type=\tw@
    {\descriptionlabel{#1}}
  \hspace{0pt}%
\else
  \ifnum\enit@type=\z@
      \hspace{-15pt} \refstepcounter{\@listctr}\fi
    \quad\@itemlabel\hspace{0pt}%
\fi}
\makeatother
\DeclareMathAlphabet{\mathdutchcal}{U}{dutchcal}{m}{n}

\def\fnote#1#2{\begingroup\def\thefootnote{#1}\footnote{#2}
     \addtocounter{footnote}{-1}\endgroup}
  
\tikzset{
    partial ellipse/.style args={#1:#2:#3}{
        insert path={+ (#1:#3) arc (#1:#2:#3)}
    }
}

\tikzset{cross/.style={cross out, draw=black, fill=none, minimum size=2*(#1-\pgflinewidth), inner sep=0pt, outer sep=0pt}, cross/.default={2pt}}

\tikzset{
	pics/torus/.style n args={3}{
		code = {
			\providecolor{pgffillcolor}{rgb}{1,1,1}
			\begin{scope}[
				yscale=cos(#3),
				outer torus/.style = {draw,line width/.expanded={\the\dimexpr2\pgflinewidth+#2*2},line join=round},
				inner torus/.style = {draw=pgffillcolor,line width={#2*2}}
				]
				\draw[outer torus] circle(#1);\draw[inner torus] circle(#1);
				\draw[outer torus] (180:#1) arc (180:360:#1);\draw[inner torus,line cap=round] (180:#1) arc (180:360:#1);
			\end{scope}
		}
	}
}

\tikzset{
	pics/hole/.style n args={2}{
		code = {
			\draw[fill=white] (0,0) arc(120:60:#1 and #2)  arc(-60:-120:#1 and #2);
            \draw (0,0) arc(-120:-130:#1 and #2) (#1,0) arc(-60:-50:#1 and #2);
		}
	}
}

\makeatletter
\newcommand*{\itemequation}[3][]{%
  \item
  \begingroup
    \refstepcounter{equation}%
    \ifx\\#1\\%
    \else  
      \label{#1}%
    \fi
    \sbox0{#2}%
    \sbox2{$\displaystyle#3\m@th$}%
    \sbox4{\@eqnnum}%
    \dimen@=.5\dimexpr\linewidth-\wd2\relax
    \ifcase
        \ifdim\wd0>\dimen@
          \z@
        \else
          \ifdim\wd4>\dimen@
            \z@
          \else 
            \@ne
          \fi 
        \fi
      \@latex@warning{Equation is too large}%
    \fi
    \noindent   
    \rlap{\copy0}%
    \rlap{\hbox to \linewidth{\hfill\copy2\hfill}}%
    \hbox to \linewidth{\hfill\copy4}%
    \hspace{0pt}
  \endgroup
  \ignorespaces 
}
\makeatother  

\hypersetup{
	pdftitle={The Standard Model from String Theory},    
	pdfauthor={\textcopyright\ Fernando Marchesano, Gary Shiu, Timo Weigand},     
	pdfsubject={hep-th},   
	pdfcreator={pdfLaTex},   
	pdfproducer={LaTex}, 
	pdfkeywords={},
	colorlinks=true
  ,urlcolor=blue
  ,anchorcolor=blue
  ,citecolor=blue
  ,filecolor=blue
  ,linkcolor=blue
  ,menucolor=blue
  ,linktocpage=true
}

\allowdisplaybreaks[1] 

\crefname{figure}{Figure}{Figures}
\crefname{table}{Table}{Tables}
\crefname{definition}{Definition}{Definitions}
\crefname{proposition}{Proposition}{Propositions}
\crefname{claim}{Claim}{Claims}
\crefname{conjecture}{Conjecture}{Conjectures}

\renewenvironment{abstract}
 {\small
  \begin{center}
  \bfseries \abstractname\vspace{-.5em}\vspace{0pt}
  \end{center}
  \list{}{%
    \setlength{\leftmargin}{4mm}
    \setlength{\rightmargin}{\leftmargin}%
  }%
  \item\relax}
 {\endlist}

\begin{document}
	\pagestyle{plain}

  \setlength{\sboxrule}{0.5em}
  \setlength{\sboxsep}{1.5em} 
  \setlength{\sdim}{7pt}

	\makeatletter
	\@addtoreset{equation}{section}
	\makeatother
	\renewcommand{\theequation}{\thesection.\arabic{equation}}
	\pagestyle{empty}
\rightline{IFT-UAM/CSIC-24-01}
 \rightline{ZMP-HH/24-01}
\vspace{1.0cm}

\begin{center}
{\large \bf
The Standard Model from String Theory: \\ What Have We Learned?

} 

\vskip 9 mm

Fernando Marchesano${}^1$, Gary Shiu${}^2$, Timo Weigand${}^{3,4}$ 

\vskip 9 mm

\small ${}^{1}$\textit{Instituto de F\'{\i}sica Te\'orica UAM-CSIC, c/ Nicol\'as Cabrera 13-15, 28049 Madrid, Spain} 

\vspace{2mm}

\small ${}^{2}$\textit{Department of Physics, University of Wisconsin-Madison, Madison, WI 53706, USA} 

\vspace{2mm}

\small ${}^{3}$\textit{II. Institut f\"ur Theoretische Physik, Universit\"at Hamburg, Luruper Chaussee 149,\\ 22607 Hamburg, Germany} 

\vspace{2mm}

\small ${}^{4}$\textit{Zentrum f\"ur Mathematische Physik, Universit\"at Hamburg, Bundesstrasse 55, \\ 20146 Hamburg, Germany  }   \\[3 mm]

\fnote{}{\hspace{-0.75cm} fernando.marchesano at csic.es, \\ shiu at physics.wisc.edu, \\  timo.weigand at desy.de}

\end{center}


\begin{abstract}

Amidst all candidates of physics beyond the Standard Model, string theory provides a unique proposal for incorporating gauge and gravitational interactions. In string theory, a four-dimensional theory that unifies quantum mechanics and gravity is obtained automatically if one posits that the additional dimensions predicted by the theory are small and curled up, a concept known as compactification. The gauge sector of the theory is specified by the topology and geometry of the extra dimensions, and the challenge is to reproduce all of the features of the Standard Model of Particle Physics from them. We review the state-of-the-art in reproducing the Standard Model from string compactifications, together with the lessons drawn from this fascinating quest. We describe novel scenarios and mechanisms that string theory  provides to address some of the Standard Model puzzles, as well as the most frequent signatures of new physics that could be detected in future experiments. We finally comment on recent developments that connect, in a rather unexpected way, the Standard Model with Quantum Gravity, and that may change our field theory notion of naturalness.

\end{abstract}

{\it Invited Contribution for Annual Review of Nuclear and Particle Science} \\
\indent https://doi.org/10.1146/annurev-nucl-102622-01223

	\newpage
	\setcounter{page}{1}
	\pagestyle{plain}
	\renewcommand{\thefootnote}{\arabic{footnote}}
	\setcounter{footnote}{0}
	
	\tableofcontents

\section{INTRODUCTION}
Since its inception, string theory has remained deeply intertwined with the field of particle physics. 
As its theoretical framework evolved, it emerged from a theory of the strong interaction to become a quantum theory of gravity.
Over the years, the synergy between string theory and particle physics has deepened, driven by the construction of increasingly realistic models and 
a better understanding of their stringy features. 
The unification of fundamental forces necessitates 
an ultraviolet-complete theory that simultaneously incorporates
both quantum gravity and chiral gauge theories.  The desire to find a consistent overarching framework that unifies particle physics with gravity was the
backdrop for heterotic strings \cite{Gross:1984dd} and Calabi-Yau
compactifications \cite{Candelas:1985en}. 
Since then, the symbiosis between string theory and particle physics has undergone a significant transformation.  
 This review provides a brief survey of this
important research area, often known as ``string phenomenology'',
highlighting some recent developments and the opportunities they present. More than a decade has passed since an  authoritative textbook on this subject has been written \cite{Ibanez:2012zz}, and this article is intended as an update of the new developments (see also \cite{Cvetic:2022fnv,Marchesano:2022qbx} for recent reviews).

Some of the central questions in this area of research are:

\noindent
{\bf 1) How well can we match observable particle physics with string constructions?}
This has been the driving question in string phenomenology for the past decades, leading to a variety of different approaches of string constructions that capture the gauge group, matter content, and other features of the Standard Model of Particle Physics (SM). 
Despite substantial progress, hand in hand with a considerably deepened understanding of the mathematics underlying string compactifications, there are still many challenges in constructing
 fully realistic string vacua that contain all its observed features including Yukawa couplings, the Higgs sector,
supersymmetry breaking, the
detailed structure of the SM parameters, and the observed cosmological constant.

\noindent
{\bf 2) What can string theory rule out?}  While 
the enormous Landscape may naively suggest 
that ``anything goes'' in
string theory, it should be noted that UV consistency with quantum gravity
imposes fairly stringent constraints on the set of allowed models, including their particle spectra and interactions.
 Finding the principles behind these quantum gravity constraints is the motivation behind the Swampland program, but the question of what is (im)possible in string theory has since the beginning been driving the field of string phenomenology.
Low energy constraints such as anomaly cancellation can be rephrased in terms of geometric constraints of the internal space.
Exploring other, perhaps more subtle, UV constraints has been a vibrant research direction in recent years.

\noindent
{\bf 3) What particle physics features of string vacua are typical?} 
 The set of 
possible consistent string
vacuum solutions is enormous, but it is believed to be finite, a feature that distinguishes string theory from other approaches to physics beyond the Standard Model (BSM).
Without a precise formulation of the measure on the Landscape, we can quantify the likelihood of certain aspects being realized by a naive counting of vacua.
Features such as hidden gauge and matter
sectors, axion fields, etc., are considered as typical in ``most'' string vacua.
This is because these characteristics can be found without extensive fine-tuning --
quantified in terms of numbers of continuous moduli or discrete
parameters like fluxes that must take special values. This notion of typicality often goes under the umbrella of ``stringy naturalness".

In addition, string theory has been a 
continuous wellspring of ideas
for particle phenomenology. 
 Arguably the thorniest conceptual problem in particle physics today is the Hierarchy Problem, i.e. the question why there is a huge disparity between the weak
scale and the Planck scale.
Much of the efforts in BSM physics over the past few decades has been directed toward addressing this question.
As reviewed below, string
theory has shown to be resourceful not only in realizing previously proposed BSM
scenarios but suggesting new ones. 
An even more vexing naturalness problem involving gravity is the smallness of the cosmological constant. It is conceivable that string theory, being a consistent quantum theory of gravity, can shed light on this puzzle.

Many of the unsolved problems in particle physics and cosmology are deeply
intertwined. We focus on particle physics aspects of string theory in this
review, but we refer readers interested in cosmological aspects to a recent survey on this topic \cite{Flauger:2022hie}.


\section{THE STANDARD MODEL FROM STRING THEORY}

String theory provides a rich framework by which one can obtain 4d effective field theories (EFTs) that resemble, to substantial accuracy, the Standard Model of Particle Physics or its extensions. Among the variety of schemes that have been developed to this end, those from which most lessons have been drawn correspond to geometric compactifications of string theory where the six extra dimensions are large compared to the string length $\ell_s$. The main advantage of these constructions is that they allow us to develop a dictionary between field theory quantities and geometry \cite{Green:1987mn,Becker:2006dvp,Blumenhagen:2006ci,Blumenhagen:2013fgp,Baumann:2014nda,Tomasiello:2022dwe,Marchesano:2022qbx}. As a result SM and BSM features, questions and puzzles can be reformulated in terms of the geometry of extra dimensions. This property has heavily influenced the strategy to build realistic High Energy Physics models  from string theory, as well as the development of most string-inspired scenarios. It is for this reason that our discussion mostly focuses on large-volume compactifications to four dimensions. For purely conformal field theoretic constructions, typically available only at special points of the moduli space, see  \cite{Lerche:1986cx, Dijkstra:2004cc, Faraggi:2006qa} and references therein.

In most of the constructions supersymmetry is recovered at the compactification scale, defined as $M_c = l^{-1}_c$, where $l_c$ is a typical length scale on the compact space. Depending on the details of the model, $M_c$ varies within the range $10^{10} - 10^{17}$ GeV. Assuming supersymmetry at the compactification scale ensures stability, and in particular the absence of dangerous tachyons.\footnote{\label{footnote1} We will not discuss here compactifications of non-supersymmetric string theories and recent attempts in the literature, for instance \cite{Abel:2015oxa}, to overcome the challenges they face such as tachyonic instabilities and a string-scale cosmological constant.} For this reason we will focus on attempts to engineer the appearance of the Minimal Supersymmetric Standard Model (MSSM) as a 4-dimensional Effective Field Theory (EFT) at low energies.
 This requisite restricts the shape of the extra dimensions and simplifies the construction of vacua, by reducing second order equations of motion to first order ones, both in the gravity and gauge sectors of the theory. The simplest of these constructions are based on manifolds of special holonomy, like Calabi--Yau threefolds (CY$_3$). These are not the most general backgrounds that one may consider to solve the higher-dimensional equations of motion, but they form a large enough set to display all of the possibilities of SM model building discovered so far, and the lessons drawn from them.\footnote{A larger set of backgrounds that solve the higher-dimensional equations of motion is given by $G$-structure manifolds \cite{Gra_a_2006,Tomasiello:2022dwe}. From a 4d EFT viewpoint, this larger class of backgrounds can be thought of as implementing potentials for light fields and supersymmetry-breaking terms. A full treatment of this larger category of compactifications is beyond the scope of the present text.} 

\begin{figure}[h]
\begin{center}\includegraphics[width=4.25in]{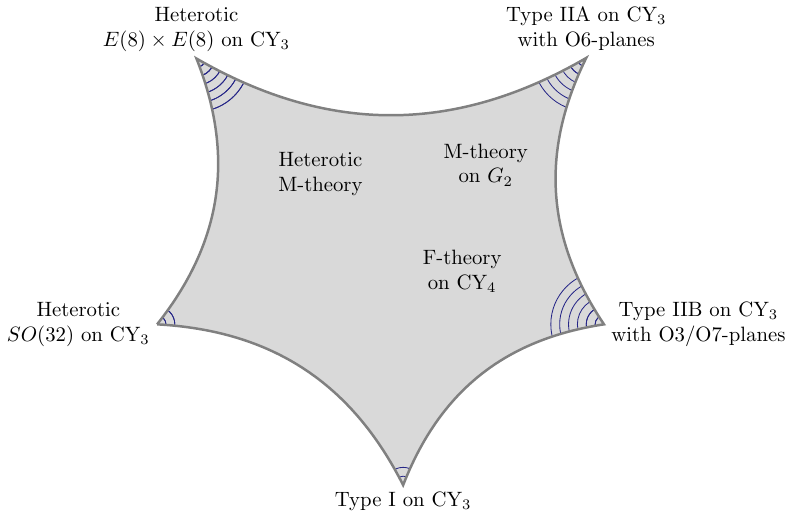}\end{center}
\caption{Regions of string theory parameter space hosting families of SM-like constructions. Corners represent constructions at weak string coupling  and the interior their strong coupling version.} 
\label{Mtheory}
\end{figure}

As of today, the different string  models that  realize the SM can be divided in two classes: those at weak and strong string coupling. Weakly coupled constructions are based on the five supersymmetric perturbative string theories formulated in 10d. The six extra dimensions correspond to a manifold that is either a compact CY$_3$ or its quotient  by a discrete symmetry of the theory, namely an orbifold or orientifold quotient. The families of constructions that are more developed in searching for the SM are illustrated in \textbf{Figure \ref{Mtheory}}. They are: 

\begin{itemize}

\item[-] $E_8 \times E_8$ heterotic string theory on a toroidal orbifold.

\item[-] $E_8 \times E_8$ heterotic string theory on a smooth CY$_3$.

\item[-] Type IIB string theory on a CY$_3$ orientifold with O3/O7-planes, and with D3-branes at singularities.	

\item[-] Type IIB string theory on a CY$_3$ orientifold with O3/O7-planes, and with intersecting D7-branes.	

\item[-] Type IIA string theory on a CY$_3$ orientifold with O6-planes, and with intersecting D6-branes. 

\end{itemize} 

Further families of perturbative constructions involve $SO(32)$ heterotic and Type I string
theory on either a CY$_3$ or a toroidal orbifold, but their features are very similar to those
of $E_8 \times E_8$ heterotic models.
 In general, all these different families of compactifications are related to each other by the web of dualities. However, it happens that some SM features are more natural in one particular frame. 

In some cases, the behavior of string theory at strong coupling is understood through string dualities.  This gives additional families of constructions that take advantage of the strong coupling regime.  The most developed examples are:

\begin{itemize}

\item[-] F-theory on a CY$_4$.

\item[-] M-theory on a $G_2$ manifold.

\item[-] Heterotic M-theory on CY$_3 \times S^1/\mathbb{Z}_2$.	

\end{itemize} 

Among them, F-theory constructions are the most sophisticated ones in several aspects, and have been found to provide a rich enough framework to incorporate many of the phenomenologically appealing features that were previously found in the different perturbative constructions. For this reason, they are usually regarded as a representative sector of the (large-volume) string theory Landscape. 

We will not attempt to describe the particularities of each of these families of string theory constructions, whose details can be found in \cite{Blumenhagen:2006ci,Ibanez:2012zz}. Instead, we focus on highlighting their common features, as well as the insights acquired when trying to reproduce the SM from them. 

As already mentioned, string theory compactifications provide a fruitful connection between the geometry of extra dimensions and the physics of EFTs. An early lesson that one obtains upon exploring this link is that the more an EFT quantity is protected against quantum corrections, the simpler is its description in geometric terms. In the context of the compactifications described above, protection mechanisms involve both gauge invariance and supersymmetry at or below the compactification scale. Typically, discrete EFT data protected by gauge invariance are described in terms of the topology of the extra dimensions, while quantities protected by supersymmetry enjoy a simple description in terms of differential and algebraic geometry. Examples in the former category are the gauge group and the chiral spectrum of the EFT, and examples of the latter are protected couplings. For this reason, we organize our discussion in terms of the string theory realization of each of these quantities when building the SM.

\subsection{Gauge Group} \label{sec:Gaugegroup}

Gauge degrees of freedom can have two different origins in perturbative string theory. In the perturbative heterotic string, both the gauge and the gravitational degrees of freedom originate from the excitations of closed strings propagating in ten-dimensional spacetime.
 The two possible types of gauge algebras in supersymmetric ten-dimensional heterotic strings, $G_{\rm 10d} = E_8 \times E_8$ or $SO(32)$,
arise from global symmetry currents
propagating along the closed string worldsheets.
When $d$ of the extra dimensions are compactified on a toroidal background, such as a toroidal orbifold \cite{Dixon:1985jw, Dixon:1986jc},
this ten-dimensional gauge sector is augmented by an additional Kaluza-Klein gauge sector, which generically contributes an extra $U(1)^d$ gauge algebra and which at special points in the moduli space enhances to a non-Abelian algebra of rank $d$. This leads to a perturbative gauge algebra $G_{\rm het}$ of maximal possible rank $22$ for compactifications to four dimensions, into which
the Standard Model gauge algebra can be embedded.
This gauge algebra can be broken by means of non-trivial gauge backgrounds, i.e. vacuum expectation values for the internal polarizations of the gauge fields. In toroidal heterotic compactifications this is achieved by specifying internal Wilson lines, which together with the geometric action of the orbifold partially break the original gauge algebra. In compactifications on a smooth Calabi-Yau threefold $X$, the internal gauge background corresponds to a quantized vacuum expectation for the field strength of a non-Abelian or an Abelian subgroup $H \subset G_{\rm het}$. Mathematically, such a gauge background is encoded in the curvature of a vector bundle $V$ with structure group $H$, and the unbroken gauge group is the commutant of $H$ within $G_{\rm het}$. When $H$ contains Abelian factors, some or all of the commuting $U(1)$s  develop a mass through a St\"uckelberg coupling to an axion \cite{Blumenhagen:2005ga}. If $X$ admits a discrete first fundamental group, the gauge group can be broken further by discrete Wilson lines, i.e. flat gauge backgrounds along torsional one-cycles on $X$ \cite{Green:1987mn}. 
 For example the SM gauge group can be obtained in this way by first breaking 
 $E_8 \to {\rm Spin(}10)$ or $SU(5)$ by a vector bundle of structure group $H= SU(4)$ or $H=SU(5)$, respectively, and then further invoking discrete Wilson lines. 

A remarkable result of string theory is that, alternatively, gauge interactions can be localized in space-time. 
 The simplest realization of this idea is given by D-branes, lower-dimensional hypersurfaces in the 10-dimensional space. A D$p$-brane extends along $p$ spatial directions plus time.
Such objects are necessary to reproduce the SM in Type II compactifications \cite{Dixon:1987yp}. In the SM realization by D-branes, each gauge factor comes from a $U(N)$ group localized in a slice of the extra dimensions, more precisely in a $(p-3)$-submanifold $\Pi$ of a compact six-dimensional manifold $X$ where $N$ D$p$-branes are wrapped, or bound states of these objects. As a consequence, each of these $U(N)$ factors is associated to a homology class of $X$, which is the topological information that describes the D$p$-brane charge from a 10d perspective. 

A Type II orientifold model is specified by the compact manifold $X$, the orientifold projection and the resulting fixed planes (O-planes), and finally the D-brane content. The O-plane and D-brane sectors define a set of homology classes of $X$, which are the topological data that determine the 4d gauge group and the chiral spectrum. Thanks to this simple description one can perform a general analysis of 4d chiral gauge anomalies. One finds that the cancellation of the total D-brane charge (a condition known as RR tadpole cancellation) and a 4d generalization \cite{Sagnotti:1992qw} of the Green-Schwarz mechanism guarantees the cancellation of chiral anomalies. As a byproduct of the Green-Schwarz mechanism several $U(1)$ symmetries are rendered massive via a St\"uckelberg mechanism, and are perceived at low energies as global $U(1)$'s that are only broken at the non-perturbative level. This is how one is able to recover the exact SM gauge group from an initial $U(3) \times U(2) \times U(1)^n$ D-brane group, together with (perturbative) SM global symmetries like $U(1)_{B-L}$ \cite{Ibanez:2001nd}.

Most of these features generalize beyond the heterotic and D-brane setups. In general, the 4d gauge group is associated with a set of  objects in the extra dimensions, like submanifolds and gauge bundles on them. These objects have a charge and a tension from a 10d viewpoint, and both must cancel to ensure anomaly cancellation and stability of the gauge sector of the theory. This statement applies to the full gauge sector, which in the most realistic models involves a SM plus a hidden gauge sector. 
F-theory \cite{Vafa:1996xn} compactifications with $[p,q]$ 7-branes, as reviewed e.g. in \cite{Weigand:2018rez,Cvetic:2018bni}, form the currently best understood example of such non-perturbative generalizations. 
In these constructions, the available gauge algebras include the full list of ADE type Lie algebras as well as their non-simply laced cousins. What makes this class of constructions particularly interesting for model building is that F-theory combines the possibility of engineering gauge algebras based on exceptional groups $E_n$ with $n=6,7,8$, known from heterotic strings, with the idea of localization of gauge degrees of freedom, as in perturbative Type II orientifolds. This has been studied systematically beginning with \cite{BeasleyHeckmanVafaI,BeasleyHeckmanVafaII, DonagiWijnholtGUTs, DonagiWijnholtModelBuilding}. 
 Such constructions are dual to compactifications of M-theory on (singular) special holonomy spaces, where the non-Abelian gauge degrees of freedom result from M2-branes wrapping vanishing 2-cycles.

An important property of string vacua is that for a given compact manifold $X$, the rank of the gauge groups is constrained by stability. In specific setups one obtains general constraints, like the upper bound for the rank of 22 in perturbative heterotic compactifications mentioned already \cite{Ibanez:2012zz}. Non-perturbative constructions like F-theory models do not obey such a simple bound, and models with a gauge group of much larger rank have been constructed \cite{Candelas:1997eh,Candelas:1997pq}, cf Section \ref{sec_darksector}.

Another interesting lesson is how string theory models realize gauge coupling unification. To leading order, the magnitude of a 4d gauge coupling is specified by how strongly the gauge interaction is diluted in the extra dimensions. In the case of a gauge interaction from a D-brane, this depends on the string coupling $g_s$ and on the volume ${\cal V}(\Pi_a)$ (in units of the string length $\ell_s$) of the submanifold $\Pi_a$ of $X$ that it wraps, leading to 
\begin{equation}
\frac{1}{g_a^2} = \frac{2 \pi}{g_s} {\cal V}(\Pi_a) \,,
\end{equation}
up to subleading corrections. For heterotic compactifications the relevant volume is that of the entire internal space $X$ and the result is suppressed by $g_s^2$ rather that $g_s$ since the gauge degrees of freedom descend from the closed string sector.

This gives the compactification-scale boundary condition for the usual renormalization group analysis of couplings in the 4d EFT.  The size and shape of $X$ is described by the vacuum expectation value of massless scalar fields called compactification moduli. The compactification moduli will  translate into SM-singlet fields in the 4d EFT. These have a variety of influences on string constructions, and we will discuss more of their implications below. What is important for us here is that the boundary conditions for different gauge couplings can depend in different ways on these moduli. This together with the fact that the SM can be built from D-branes of different sizes and dimensions shows that gauge coupling unification is not a universal feature of string constructions.

Approximate gauge coupling unification is realized when all of the SM gauge interactions come from a common source. Schematically, such models start from an initial large gauge group $G$, like $SU(5)$, Spin$(10)$ or $E_6$, which is then endowed with a gauge-group breaking mechanism that yields the SM gauge group. The new mechanisms that string theory provides for gauge-group breaking compared to standard BSM setups are based on the the fact that, microscopically, $G$ arises from a higher-dimensional gauge field. Therefore a non-trivial profile for such a gauge field in the extra dimensions can break $G$ to the SM group, just like for the heterotic gauge groups $G_{\rm het}$ discussed above.  
More details on setups with gauge coupling unification  will be provided in Section \ref{subsec_GaugeUni}.

\subsection{Chiral Spectrum}

A general lesson from building 4d EFTs from string theory is that chiral fermions are localized in the extra dimensions. This feature is particularly manifest in Type IIA intersecting D6-brane models. In these models chiral fermions are localized at the points of intersection of two three-cycles of the compact manifold $X$ \cite{Blumenhagen:2005mu}. In heterotic compactifications on a smooth CY$_3$ the localization is less dramatic, but it also occurs because a chiral fermion is represented by a non-trivial wavefunction in the extra dimensions, whose profile peaks at some particular point. In intersecting D7-branes and F-theory models, chiral fermion localization is a combination of these two mechanisms. As we will discuss in the next section, chiral fermion localization is crucial to obtain a realistic pattern of Yukawa couplings in this class of constructions.

The number of SM fermions of each type in the 4d EFT is given by the net number of chiral fermion states (that is, up to vector-like pairs) within the given gauge group representation. In string theory models, this net chirality is always related to some topological index of the compactification. In the case of intersecting D6-branes, it is simply the signed intersection number of two three-cycles. In heterotic compactifications on smooth CY$_3$ with internal gauge bundles, it corresponds to the index of the corresponding Dirac operator \cite{Green:1987mn,Ibanez:2012zz}. This simple description of the chiral spectrum has led to a general modular strategy to find string theory models that contain the SM gauge group and chiral spectrum and are absent of chiral exotics \cite{Marchesano:2022qbx}. The strategy works particularly well for chiral theories with only 2-index tensor representations in the gauge sector, as it happens in D-brane models, and it is equally well suited for smooth Calabi--Yau compactifications, D-branes at singularities \cite{Maharana:2012tu} and Gepner orientifold models \cite{Dijkstra:2004ym}. It can also be adapted to other constructions, like F-theory models \cite{Lin:2014qga} and heterotic compactifications with line bundles \cite{Blumenhagen:2005ga}. The description of massless vector-like pairs is in general more involved, as it requires the knowledge of finer geometric data, see e.g. \cite{Bies:2017fam} in the context of F-theory. However, in concrete models these vector-like fermions must acquire mass. Several techniques are available for this. These include giving a vev for SM singlets or switching on discrete Wilson lines, internal gauge fluxes or background fluxes. 

All this progress has led to more and more sophisticated string theory realizations of the SM gauge group and chiral field content \cite{Cvetic:2001nr,Cvetic:2001tj,Marchesano:2004xz, Raghuram:2019efb,Cicoli:2021dhg}, without any SM chiral exotic matter \cite{Bouchard:2005ag,Braun:2005nv}. The state-of-the-art model building has made possible systematic scans recovering large numbers of realistic compactifications from them \cite{Dijkstra:2004cc,Lebedev:2006kn,Gmeiner:2007zz,Anderson:2012yf,Cvetic:2019gnh}.

Certain compactification schemes like D-branes at singularities have an upper bound of three SM families, which comes from the number of (complex) extra dimensions \cite{Maharana:2012tu}. This bound is, however, not observed in other setups, and at present there is no compelling argument of why three generations should be preferred from a string theory viewpoint. What instead can be thought of as a universal feature of string theory models is that the light fields generically transform in small  representations of the gauge group, compared to the existing possibilities in field theory. This is clear in D-brane models, where the open string spectrum always yields at most 2-index representations of the gauge group. Note that this is just enough to accommodate the representations occurring in the SM. In more general M and F-theory compactifications higher representations are possible \cite{Klevers:2017aku}, but non-generic \cite{Taylor:2019ots}, with the current record held by \cite{Li:2022vfj}. Establishing bounds on the maximal possible charge is an open problem \cite{Raghuram:2018hjn}.

\subsection{Yukawa Couplings}

The physical Yukawa couplings of string compactifications are described by the standard 4d supergravity expression 
\begin{equation}
Y_{ijk}  = (K_{i\bar{i}} K_{j\bar{j}} K_{k\bar{k}})^{-1/2} e^{K/2} W_{ijk} \, .
\label{yukis}
\end{equation}
with $K$ the K\"ahler potential of the 4d supergravity theory, $W_{ijk} \Phi^i \Phi^j \Phi^k$ a cubic term in the superpotential involving three chiral fields, and $K_{i\bar{i}} = \partial_{\Phi^i} \bar{\partial}_{\bar{\Phi}^i} K$  their K\"ahler metrics. Just like gauge couplings, each of these quantities depends on compactification moduli: $W_{ijk}$ is a holomorphic and $K$ a real function of the 4d fields. 
Even computing the holomorphic piece  
$W_{ijk}$ for compactifications on general Calabi-Yau varieties requires advanced mathematical techniques (see \cite{Lerche:2018gvn} and references therein).
In general, the K\"ahler metrics for chiral fields is even more difficult to obtain beyond simple setups. A general expression for the K\"ahler potential including the chiral fields of the compactification remains as an open problem. In fact, it is oftentimes simpler to directly compute the physical Yukawa couplings instead of each the quantities in the rhs of (\ref{yukis}), a strategy followed for instance in \cite{Cremades:2003qj,Cremades:2004wa}. 

The basic Yukawa structure that arises from string compactifications is the following. If the coupling results from a point interaction, like for D-branes at singularities or three D-branes intersecting at a point, then the Yukawa is of order one. If the Yukawa is generated by an overlap of three wavefunctions in the extra dimensions, then there can be a suppression due to the separation of the wavefunction peaks from each other and from wavefunction normalization. With these two features and assuming a minimal Higgs sector, one may try to engineer the observed SM flavor hierarchies. Additional ingredients, elaborated on in Section \ref{sec:selectionrules}, which generate Yukawa textures involve global $U(1)$ symmetries that forbid certain Yukawa couplings at the perturbative level \cite{Ibanez:2008my}, as well as flavor-dependent discrete gauge symmetries \cite{Nilles:2012cy,Berasaluce-Gonzalez:2012abm,Marchesano:2013ega}. 

Despite the progress made in computing and understanding Yukawa couplings, it is fair to say that currently there is no general scheme to reproduce flavor hierarchies. It is however important to stress that string models have been constructed that reproduce the observed SM flavor hierarchies. 
To name one example, a context in which this has been achieved are F-theory GUT models, discussed in more detail in Section \ref{subsec_GaugeUni}.
The starting point is a Yukawa matrix that at tree-level is of rank one, as can be easily engineered because this rank is a topological quantity \cite{Cecotti:2009zf}. The rank is increased to three when non-perturbative effects from a hidden sector are included \cite{Marchesano:2009rz}, naturally leading to one heavy family of quarks and leptons. In the framework of local F-theory GUTs this scenario can be translated into precise computations \cite{Cecotti:2010bp}, which result in realistic fermion masses and mixings for large parameter regions of the MSSM \cite{Font:2012wq,Font:2013ida,Marchesano:2015dfa,Carta:2015eoh}.



\section{STRING THEORY SCENARIOS}

A central goal of string model building is to extract and understand which mechanisms exist in string theory to overcome long-standing challenges of particle physics phenomenology. This question has two aspects:
First, can known field theoretic mechanisms be realized in string theory and if so do they arise naturally? Second, are there new mechanisms in string theory 
that go beyond purely four-dimensional field theoretic approaches?

\subsection{Local versus Global Models} \label{subsec_LocalGlobal}

The search for general such mechanisms rather than explicit vacua is particularly 
beneficial when it is possible to isolate a specific question in particle physics from other aspects of the string model. 
More concretely, gauge degrees of freedom may localize in certain regions of the compactification space $X$ and hence appear, at least at first sight, to be relatively little sensitive to the gravitational degrees of freedom which are associated with global properties of the compactification space.
Examples of this are localization of the gauge sector on D-branes wrapping shrinkable cycles on $X$ - the extreme case corresponding to branes at singularities - or geometric engineering via isolated singularities in M-theory. 

This has lead to the idea of local string model building, where in a first  approach the gauge sector is analyzed by itself, while the coupling to gravitational degrees of freedom or other hidden sectors is achieved at a later step \cite{Aldazabal:2000sa}. 

Specifically, D-branes placed at singularities of Calabi-Yau threefolds \cite{Douglas:1996sw} give rise to gauge sectors and charged chiral matter in such local models which can get very close to the Standard Model. Promising classes of such singularities are orbifold-type singularities \cite{Aldazabal:2000sa,Berenstein:2001nk}, toric singularities (e.g. \cite{Krippendorf:2010hj} and references therein) or more general del Pezzo singularities such as those analyzed in \cite{Verlinde:2005jr,Wijnholt:2002qz, Buican:2006sn,Dolan:2011qu}. 
An account of progress in the direction of local models has been given in \cite{Quevedo:2014xia}.
Questions amenable to purely local techniques include the non-Abelian part of the gauge algebra,
the associated charged chiral spectrum,
and the leading contributions to the Yukawa couplings
and the gauge couplings.

This set of questions is to be contrasted with the properties of a string vacuum that are sensitive to the global properties of the model, i.e. to the embedding of the localized degrees of freedom into the compactification manifold.
In fact, one of the hallmarks of string theory is that it generically describes gauge and gravitational degrees of freedom not as being in isolation. Hence special care must be taken in
any attempt to separate (certain) particle physics aspects from the quest of coupling the theory back to gravity. As a general lesson, local mechanisms oftentimes require substantial modifications once gravitational effects are taken into account.  This applies, in particular, to the sector of Abelian continuous  \cite{Buican:2006sn} or of discrete gauge symmetries, to which we will return in Section \ref{sec:selectionrules}, to the perturbative \cite{Conlon:2009kt, Conlon:2010xb} or non-perturbative corrections to the Yukawa couplings and gauge kinetic terms,
and to all
questions related to 
moduli stabilization, supersymmetry breaking,
the stabilization of 
physical scales such as the gauge coupling unification or supersymmetry breaking scales, as well as the cosmological history, 
including the evolution of the early universe and late time acceleration.
The embedding of (semi-)realistic gauge sectors on D-branes probing local singularities into fully-fledged compact models has been advanced in \cite{Cicoli:2012vw,Cicoli:2013cha,Cicoli:2013mpa}.

\subsection{Gauge Coupling Unification}
\label{subsec_GaugeUni}

Gauge coupling unification is most natural if
the three SM gauge group factors share a common group theoretic origin in the sense of a Grand Unified Theory (GUT).
A key challenge in GUT model building is then the breaking of the GUT group $G$ to the SM gauge group $SU(3) \times SU(2) \times U(1)$.
In purely field theoretic constructions, this is achieved by a Higgs mechanism. Apart from the question of how to generate the potential of a GUT Higgs dynamically, this approach is inflicted with phenomenological challenges such as the notorious doublet-triplet splitting problem in $SU(5)$ GUTs.
It is therefore an interesting question for model building if string theory provides new mechanisms to address these challenges in a GUT context.
This question comes in two parts: First, which types of GUT groups appear naturally, and second, which mechanisms for GUT breaking are available in addition to the field theoretic Higgs mechanism? Besides naturally accomodating phenomenologically appealing GUT groups, stringy GUTs offer a topological mechanism to realize GUT breaking, which is intimately tied to the higher-dimensional nature of the theory. 
Namely, such constructions admit background values for suitable components of the gauge fields along the extra dimensions, as already mentioned in Section \ref{sec:Gaugegroup}.

These general ideas are very concretely realized already in the perturbative $E_8 \times E_8$ heterotic string. A detailed overview is given in \cite{Dienes:1996du}. Focusing for concreteness on compactifications on a smooth space $X$, a gauge background on  $X$ reduces one of the ten-dimensional $E_8$ gauge factors to a GUT group such as $G= SU(5)$, Spin$(10)$ or $E_6$. 
The GUT group can then be further broken topologically to $SU(3) \times SU(2) \times U(1)$ by extra discrete gauge backgrounds. In the context of compactifications on smooth Calabi-Yau threefolds this last step requires discrete Wilson lines, as mentioned already in Section \ref{sec:Gaugegroup}.
This approach has resulted in the exact SM spectrum embedded into a GUT scenario \cite{Bouchard:2005ag,Braun:2005nv}, including the possibility of vector-like Higgs pairs.

Alternatively, the small parametric separation between the GUT scale, $M_{\rm GUT} \sim 10^{16}$ GeV, and the four-dimensional Planck scale $M_{\rm Pl} \sim 10^{19}$ GeV, can be viewed as motivation for a local approach to unification.
The starting point is a gauge sector localized on a spacetime filling brane whose internal directions extend along a shrinkable cycle on the compactification manifold. The embedding of the Standard Model into a GUT furthermore turns out to be particularly natural in a framework that admits exceptional gauge algebras. Taken together, both features -- localization of gauge degrees of freedom on branes and appearance of exceptional gauge degrees of freedom -- point to the idea of F-theory GUTs initiated in \cite{BeasleyHeckmanVafaI,BeasleyHeckmanVafaII,DonagiWijnholtGUTs,DonagiWijnholtModelBuilding} (see \cite{Weigand:2010wm,Weigand:2018rez} for more detailed reviews).

\vspace{.25cm}
{\bf Example: SU(5) F-theory GUTs}
 Particular attention has been devoted to the GUT scenario in which an $SU(5)$ gauge group  is engineered on stack of 7-branes.\footnote{For constructions based on gauge groups Spin$(10)$, $E_6$ or $E_7$ we refer to the discussion and references in \cite{Li:2023dya}.} These wrap a holomorphic four-cycle ${\bf S}$ on the compactification space $X_3$, which in F-theory is identified with the base  of an elliptically fibered Calabi-Yau four-fold $Y_4$. 
Within a local approach ${\bf S}$ should be a shrinkable divisor and hence a generalized del Pezzo surface even though this restriction is not strictly required. 

 Despite the increasing tension with LHC results, let us focus for simplicity on an ${\cal N}=1$ supersymmetric GUT scenario.  Apart from the $SU(5)$ gauge degrees of freedom propagating along ${\bf S}$, there must exist three generations of GUT matter in the representations ${\bf 5_m}$ and ${\bf 10}$ of SU(5). The SM Higgs sits in a vector-like pair of ${\bf 5_H} + {\bf \bar 5_H}$. All these are localized on curves where ${\bf S}$ intersects additional 7-branes. In a purely local approach, these additional branes are not specified as they extend away from the localized GUT sector. Part of explaining the coupling to gravity (to address the physics questions sensitive to global details of the model, cf. Section \ref{subsec_LocalGlobal}) is about specifying these further non-GUT branes.

The Standard Model Yukawa couplings arise from the decomposition of the allowed GUT couplings ${\bf 10 \, \bf 10 \, 5_H}$ and ${\bf 10 \, \bf \bar 5_m \, \bar 5_H}$. These couplings are generated by the overlap of the wavefunction of the localized matter modes at the points of intersection between the matter curves $C_{\bf 10}$,  $C_{\bf \bar 5_m}$ and the Higgs curve. The first coupling is responsible for the top quark Yukawa. A large coupling of this sort is  hard to obtain in Type II orientifold constructions, while F-theory incorporates it naturally.  
This is related to the local enhancement of the gauge symmetry to the exceptional group $E_6$ at these Yukawa points.

If the internal space along which the gauge field propagates admits non-trivial one-cycles, there can be topologically non-trivial Wilson line backgrounds, similar to the aforementioned GUT breaking in heterotic compactifications on non-simply connected Calabi-Yau threefolds. 
 Even in absence of such one-cycles, the background gauge field can acquire quantized background values. Geometrically such configurations correspond to non-zero field strength VEVs along two-cycles. 
Specifically, the breaking 
$SU(5) \to SU(3) \times SU(2) \times U(1)_Y$ is implemented by a non-trivial VEV for the hypercharge $U(1)_Y$ field strength along ${\bf S}$, $\langle F_Y \rangle \neq 0$.

 Typically non-trivial flux backgrounds lead to a Green-Schwarz mechanism, which in the present context would  make the $U(1)_Y$ massive. This unacceptable effect can also be circumvented thanks to the brane-localization of the GUT gauge group in the compact extra dimensions: As long as the class of $F_Y$ on $S$ is topologically trivial within $X_3$, no Green-Schwarz mass is induced \cite{Buican:2006sn,Heckman:2008qa,DonagiWijnholtGUTs}.  Chiral three-generation models on smooth compact elliptic 4-folds with 4-form fluxes (see \cite{Braun:2011zm} and references therein) of this type were first constructed in \cite{Marsano:2011hv,Krause:2011xj,Grimm:2011fx}.

 Along the way, this GUT breaking mechanism in principle allows for topological ways to solve the notorious doublet-triplet splitting problem of GUTs. A recent overview of some of the proposed constructions can be found in \cite{Krippendorf:2015kta,Marchesano:2022qbx} along with open questions related to the embedding of the particle sector into a full theory including gravity.

\subsection{Mechanisms to Generate Hierarchies}

Generating hierarchies between scales or couplings is one of the key challenges in particle physics.
In string theory, all scales measured in units of the intrinsic string scale $\ell_s$, and all dimensionless couplings are dynamical in the sense that they are determined by vacuum expectation values of scalar fields.  This paves the way for new mechanisms to generate  hierarchies. We will discuss two examples for the generation of such hierarchies which cannot be explained in purely four-dimensional quantum field theory. The first relies on extra dimensions, the second on non-perturbative gravitational dynamics. 
While both effects could in principle be discussed independently of string theory, the latter naturally accommodates them and makes them computable in UV-complete frameworks.

\subsubsection{ Large extra dimensions and warpings}

The probably most prominent mechanism to generate hierarchies in scales relies on the size of the extra dimensions \cite{Arkani_Hamed_1998,Antoniadis_1998,Dienes:1998vh,Dienes:1998vg} and the possibility of a warp factor \cite{Randall_1999}. 

The most general ansatz for a compactification that respects four-dimensional Poincar\'e invariance is a warped metric of the form
\begin{equation}
ds^2 = e^{2 A(y)} g_{\mu \nu}(x) dx^\mu
dx^\nu + G_{m n}(y) dy^m dy^n \,,
\end{equation}
where $x^{\mu}$, $\mu = 0,1,2,3$, and $y^m$, $m=4, \ldots, 9$, are coordinates along the visible and extra dimensions, respectively. The warp factor $e^{2 A(y)}$ depends only on the 
internal coordinates. Non-trivial warp factors naturally occur in flux compactifications or as a consequence of brane backreactions \cite{Verlinde_2000,Klebanov:2000hb,Giddings:2001yu}.

Just like large extra dimensions, warping has the potential to create a hierarchy between ten-dimensional Planck scale and the four-dimensional Planck scale. These are related by
\begin{equation}
M_{{\rm Pl},4}^2 = M_{{\rm Pl},10}^8  \hat {V}(X) \,,   \qquad \quad  \hat {V}(X) = \int_X e^{2A(y)} \sqrt{{\rm det}(G)} \,, 
\end{equation}
with $\hat {V}(X)$ the warped volume of the compactification space.  
For large $\hat {V}(X)$ this leads to parametric differences in $M_{\rm Pl,4}$ and $M_{\rm Pl,10}$, which acts as the fundamental Planck scale of the underlying UV theory. 
While this mechanism as such is not specifically string theoretic in nature, string theory provides  realizations of this scenario in controlled setups \cite{Giddings:2001yu}.

Combining warping effects with the localization of gauge degrees of freedom on D-branes leads to more refined patterns of scales because the couplings on the branes on which the Standard Model is located is sensitive to the warp factor evaluated at the brane location on $X$, averaged over the brane. Applied to the Higgs mass, this mechanism can alleviate the tension between the electro-weak and the four-dimensional Planck scale \cite{DeWolfe:2002nn}.

\subsubsection{Selection rules and non-perturbative effects} \label{sec:selectionrules}

In many phenomenological scenarios selection rules constrain the set of possible operators in the effective action. 
In the low-energy theory these selection rules may appear as global symmetries. In quantum gravity, however, no exact global symmetries are believed to exist \cite{Banks:1988yz, Banks:2010zn} and hence such symmetries must either be gauge symmetries at high energies or be broken. 
Symmetries broken only by subleading effects give rise to approximate selection rules, and explain why some operators in the effective theory are strongly suppressed.


 Consider an operator ${\cal O}_q$ with charge $q$ under what appears to be a global $U(1)$ symmetry at  the level of the {\it perturbative} low energy effective action. 
Here perturbative may refer either to the expansion in the string coupling $g_s$ or to the worldsheet $\ell_s^2$ expansion. 
In string theory, such $U(1)$s are always associated with dynamical Abelian gauge fields which acquire a St\"uckelberg mass term $M_{U(1)}$. Above this scale they behave as gauge symmetries while at energies below they appear as perturbative global selection rules which prevent any operator ${\cal O}_q$ of charge $q\neq 0$ from appearing in the perturbative effective action. However, an inevitable consequence of the St\"uckelberg mechanism is that non-perturbative effects break this perturbative $U(1)$ selection rule to a discrete subgroup or even completely. This effect results from the existence of instantons carrying an effective charge under
the $U(1)$ \cite{Blumenhagen:2006xt,Ibanez:2006da,Haack:2006cy,Blumenhagen:2009qh}.
An instanton of $U(1)$ charge $-q$ can give rise to an operator with $U(1)$ charge $q$ in the effective action,
\begin{equation}
S \supset \int d^4 x \,  |e^{-S_{\rm inst}}| \, {\cal O}_q \,.
\end{equation}
This leads to non-perturbatively suppressed terms in the effective action and hence serves as a mechanism to naturally generate hierarchically small couplings. For explicit examples of the resulting suppression factors in the particle physics context we refer to the review \cite{Blumenhagen:2009qh} and references therein.

The instantons in question are oftentimes referred to as stringy or exotic instantons \cite{Argurio:2007vqa}. This stresses that they are not gauge instantons; in particular their existence is not tied to strong non-Abelian gauge dynamics. 
Instead, one should view them as gravitational instantons \cite{Hebecker:2018ofv}. This shows again that the coupling to gravity has important consequences also for gauge sectors.

In Type II constructions and their orientifolds, concrete realizations of such instantons are  Euclidean D$p$-branes wrapping compact $(p+1)$-cycles $\Sigma_{p+1}$ on $X$. The instanton action $S_{\rm inst}$ controlling the suppression factor is obtained by reducing the Dirac-Born-Infeld and Chern-Simons action along $\Sigma_{p+1}$.
It takes the form
\begin{equation}
S_{\rm inst} =\frac{2\pi}{g_s} {\cal V}(\Sigma_{p+1}) + i \int_{\Sigma_{p+1}} C_{p+1} + \ldots \,,
\end{equation}
where ${\cal V}(\Sigma_{p+1})$ is the volume in string units. Integrating the Ramond-Ramond form $C_{p+1}$ over $\Sigma_{p+1}$ gives rise to an axion $a$ in the 4d EFT. Precisely this axion participates in the St\"uckelberg mechanism making the $U(1)$ massive and hence transforms under the $U(1)$ symmetry 
as $a \to  a -q$. This results in the charge of the instanton operator $e^{-S_{\rm inst}}$.
The mechanism is generic in string theory and not restricted to Type II constructions. For instance, in heterotic compactifications, stringy instantons are either worldsheet instantons (and hence non-perturbative in $\ell_s$ rather than in $g_s$) or NS5-brane instantons (which are non-perturbative in $g_s$).

Note that the operator ${\cal O}_q$ that can be generated in such a non-perturbative way can in principle be of any mass dimension. Examples of interesting suppressed operators that have been investigated include supersymmetry breaking F-terms, non-perturbative Majorana masses or $\mu$-terms, or suppressed Yukawa-type interactions, for instance to generate suppressed Dirac neutrino masses in particle physics. For details and references see \cite{Blumenhagen:2009qh}.

In the presence of an instanton of charge $|q|=1$, the Abelian selection rule is broken completely at the non-perturbative level. By contrast, if all instantons carry charges which are multiples of some integer value $k$, then only operators of charge $q = 0 \,  {\rm mod}(k)$ are generated. The $U(1)$ global symmetry (in the sense explained above) is broken to a $\mathbb Z_k$ symmetry, which is, by definition, exact at the non-perturbative level. Note that in the fully-fledged string (or quantum gravity) theory, the $\mathbb Z_k$ symmetry is really a discrete gauge symmetry \cite{Berasaluce-Gonzalez:2011gos,Camara:2011jg}, rather than a global symmetry, in agreement with the lore that all exact symmetries must be gauged in quantum gravity \cite{Banks:2010zn}.  
This way of generating a $\mathbb Z_k$ symmetry is qualitatively different from the familiar field theoretic mechanism of Higgsing a $U(1)$ symmetry by a Higgs field of charge $k$: In the latter case, the $U(1)$ symmetry is restored at finite distance in the moduli space (corresponding to vanishing Higgs field), while here the $U(1)$ is restored only at infinite distance in field space, where $|S_{\rm inst}| \to \infty$.

A natural question concerns the types of discrete 
$\mathbb Z_k$ symmetries
are possible in string theory, i.e. the maximal possible order $k$. While in field theory there is no bound on $k$, it is believed that in string theory (or more generally in quantum gravity) only a finite number of values are possible. Relatedly, one may ask which are the possible charges under an Abelian gauge group in string theory. 
Ref. \cite{Lee:2022swr} argues for a bound on the order of $\mathbb Z_k$ within six-dimensional F-theory constructions based on general probe brane arguments.

An intrinsically string theoretic type of discrete symmetries are modular symmetries, e.g. with modular symmetry group $SL(2, {\mathbb Z})$. Combined with conventional discrete symmetries these offer interesting avenues to understand the flavor hierarchies in the SM, as reviewed in \cite{Nilles:2023shk}.

The key lesson from this discussion is two-fold:
First, all questions related to global symmetries and selection rules must be analyzed in the context of a quantum gravity theory, rather than for the particle sector in isolation. Second, gravitational dynamics such as charged gravitational instantons and their concrete string theoretic realization as charged stringy instantons offer a way to create hierarchical couplings in effective theories.Also,  string theory provides a geometrical understanding of such  mechanisms.

\section{PHYSICS BEYOND THE STANDARD MODEL}

String constructions that realize the Standard Model always contain additional sectors, either gravitational or gauge, at an energy scale that could be tested in the near future. The precise field content is quite model dependent, but there are certain BSM features that are typically found when one tries to reproduce the SM from string theory. In the following we exemplify some of them that could be detected in future experiments.

\subsection{Neutrino Sector}

Right-handed neutrinos are not part of the Standard Model, but they are probably the simplest explanation for the observed spectrum of neutrino masses and mixing angles, via the see-saw mechanism. They oftentimes appear in semi-realistic string theory constructions, either because the model is based on a GUT gauge group like Spin$(10)$ or $E_6$ that contains a right-handed neutrino in its basic representation, or because it contains the perturbative symmetry  $U(1)_{B-L}$, which becomes massive through a St\"uckelberg mechanism as described in Section \ref{sec:Gaugegroup}.

The question is whether  string compactifications can generate the necessary couplings that reproduce the see-saw mechanism. This requires the generation of a Majorana neutrino mass. The potential difficulty is that, while right-handed neutrinos are not charged under the SM gauge group, they are charged under additional symmetries like $U(1)_{B-L}$, which forbid neutrino Majorana masses at the perturbative level. Nevertheless, two main mechanisms have been observed that overcome this obstruction. The first one has arisen in the context of heterotic orbifolds, see  \cite{Faraggi:1993zh,Giedt:2005vx,Buchmuller:2007zd}, which can include superpotential couplings of the form
\begin{equation}
\Lambda^{n-1}  h_{ij} N^i N^j \prod_{a=1}^n s^a \,.
\end{equation}
Here $N^i$ are the right-handed neutrinos, $\Lambda$ is either the compactification or the string scale and $s^a$ are some SM singlet fields which acquire large vevs upon moduli stabilization. 

The second mechanism arises more naturally in Type II orientifold models \cite{Blumenhagen:2006xt,Ibanez:2006da,Blumenhagen:2009qh}, and it is based on non-perturbative effects breaking the $U(1)$ symmetries that forbid Majorana masses, generating couplings of the form
\begin{equation}
M_s e^{-S_{\rm inst}} h_{ij} N^i N^j\, .
\label{Einst}
\end{equation}
Here $S_{\rm inst}$ is the classical action of an instanton that breaks, e.g., $U(1)_{B-L}$
as explained in Section \ref{sec:selectionrules}.
 A key point which we recall from there is that this can be a gravitational instanton and so, unlike for ordinary gauge theory instantons, its action is not constrained by any SM gauge coupling. Thanks to this feature one can generate Majorana masses of the required magnitude. 

In both of these setups the right-handed neutrinos clearly belong to the gauge sector of the theory, as they are highly localized in the extra dimension and charged under some extension of the SM gauge group. This does not need to be the case in general, as has been pointed out in the context of $SU(5)$ F-theory GUTs \cite{Bouchard:2009bu,Tatar:2009jk}. It this setting the right-handed neutrino candidates are complex structure moduli of the bulk \cite{Tatar:2009jk}, which gain a mass via a moduli stabilization mechanism, or even Kaluza-Klein modes \cite{Bouchard:2009bu}. Finally, in models without Majorana masses for right-handed neutrinos, there exist alternative mechanisms to generate realistic neutrino masses. In particular, instanton effects can also generate a superpotential term of the form
\begin{equation}
M_s^{-1} e^{-S_{\rm inst}} L H_u L H_u\, ,
\label{EinstW}
\end{equation}
which leads to a dimension five Weinberg operator.

\subsection{Moduli and Axions}

A generic property of compactifications of string theory is the appearance of a plethora of
light scalar fields whose masses lie at or below the compactification scale.
In addition to the dilaton present already in ten dimensions, closed string moduli arise as massless fluctuations of the metric reduced along the compactified extra six dimensions. They are in one-to-one correspondence with flat deformations of the metric. Their vacuum expectation values parametrize the size and shape of the compactification space. For example, in Type II compactifications on Calabi-Yau threefolds, the closed string moduli space factors into deformations of the K\"ahler and the complex structure. Similarly, open string moduli are associated with deformations of spacetime-filling branes along the compactification space. 
In 4d compactifications with at most ${\cal N}=1$ supersymmetry, various effects can induce a scalar potential on the moduli space, i.e. on the space of scalar fields, which may lead to non-vanishing mass terms for some or even all moduli in the minimum of the potential. Finding minima with no massless moduli is the goal of the moduli stabilization program as reviewed e.g. in \cite{Gra_a_2006}. The best studied examples include Type IIA compactifications with background fluxes stabilising both K\"ahler and complex structure moduli \cite{DeWolfe:2005uu} and Type IIB compactifications with background fluxes to stablize the complex structure moduli and quantum effects to stabilize the K\"ahler moduli \cite{Kachru:2003aw,Balasubramanian:2005zx}. Similar effects also give rise to moduli masses of other classes of compactifications, see Section \ref{sec_susybr}. 
Since the value of the moduli determine the couplings of the gauge sector, detailed questions about the size of the couplings cannot be analyzed in isolation from the moduli problem.

At least in scenarios with low-energy supersymmetry, one or several moduli tend to acquire masses around the supersymmetry breaking scale and hence are rather light. This brings challenges for cosmology, known as the cosmological moduli problem,s see for instance \cite{Quevedo:2014xia} and references therein. 

Another generic property of string compactifications is the appearance of axions \cite{Svrcek:2006yi}, at least prior to taking into consideration the effects of moduli stabilization.
Axions arise naturally by reducing the Ramond-Ramond $p$-form gauge potentials $C_p$
or the Neveu-Schwarz 2-form potential $B_2$ over appropriate cycles, for instance as $a = \int_{\Sigma_p} C_p$, where $p$ is a $p$-cycle on the compactification space. Open string axions arise from 1-form gauge potentials on D-branes, reduced over internal 1-cycles of the brane. 
In the low-energy supergravity theory, axions pair up with the moduli to form complex scalar fields.
Unlike their partners, the geometric moduli, they enjoy a perturbative shift symmetry which is inherited from  the gauge symmetry of the underlying $p$-form potential. The perturbative shift symmetry can be broken by non-perturbative effects, i.e. by worldsheet or D-brane instantons in string theory, or in the presence of background sources such as gauge fluxes or wrapped branes.
The axionic couplings in the EFT arise by dimensional reduction of the Chern-Simons couplings in a higher-dimensional effective action.
For example, a D$p$-brane has a Chern-Simons action of the form
\begin{equation}
S_{\rm CS} = \frac{2 \pi}{\ell_s^{p+1}}\int_{D_p} (\sum_k C_k) \wedge  {\rm tr} e^{\frac{i \ell_s^2}{2 \pi} F} \sqrt{\frac{\hat A(TD)}{\hat A(ND)}} \,,
\end{equation}
where $\hat{A}$ is the A-roof genus of the tangent and normal bundle to the D$p$-brane, and accounts for the couplings to the spacetime curvature. In this way the axionic fields participate  in topological couplings of the form 
\begin{equation}
\int_{\mathbb R^{1,3}} a \, {\rm tr} F\wedge F \,.
\end{equation}
Analogous  couplings appear in the heterotic string by dimensionally reducing the anomalous Green-Schwarz couplings. This may in principle give rise to an axion coupling to the strong sector as envisaged in the Peccei-Quinn solution to the strong CP problem. However, the phenomenological viability of the axion depends on its mass and axion decay constant, as reviewed for instance in \cite{Cicoli:2013ana}.  
 In a 4d ${\cal N} =1$ supersymmetric Minkowksi minimum (if they exist at all \cite{Palti:2020qlc}),
the stabilization mechanism required to remove the moduli fields from the massless spectrum gives a mass of the same size to its axionic partner. 
Approximately massless axions are therefore expected to arise rather in AdS compactifications or in non-supersymmetric minima. 
In the spirit of \cite{Arvanitaki:2009fg}, however, one can read the appearance of a QCD axion solving the strong CP problem as a constraint on a string compactification, and assuming its existence suggests the appearance of several, if not hundreds of axionic particles with interesting phenomenological properties. The reason is that a typical compactification manifold exhibits ${\cal O}(100)$ compact cycles over which the higher-dimensional $p$-form potentials can be reduced.

\subsection{MSSM and Supersymmetry Breaking}
\label{sec_susybr}

While most SM aspects described so far are independent of supersymmetry, almost all string theory models that resemble the SM display supersymmetry at the compactification scale, since otherwise they face severe stability issues (see footnote \ref{footnote1}). The source of supersymmetry breaking is the same potential used to stabilize light scalar fields, whose minima may or may not break supersymmetry spontaneously. As a result, the different supersymmetry-breaking scenarios are related to the moduli stabilization mechanisms available in each family of constructions.

A classical scheme is the one found in $E_8 \times E_8$ heterotic compactifications, where one of the $E_8$ factors hosts the SM sector and the other one is a hidden sector. A gaugino condensate in the latter will generate a scalar potential that will both stabilize moduli and break supersymmetry, which is then communicated to the visible sector via gravitational effects \cite{Derendinger:1985kk,Dine:1985rz}. The problem with this mechanism is that the generated potential has runaway directions toward a non-interacting theory \cite{Dine:1985he} and so extra ingredients must be added  to generate a more involved potential with actual minima. One possibility is to add background fluxes, but their quantization implies that supersymmetry is broken at a high scale. Alternatively, one may consider several gauge factors that yield a gaugino condensate, so that the generated superpotential involves a sum of competing exponentials, also known as a {\it racetrack scenario} \cite{Krasnikov:1987jj}. This idea can be made precise in toroidal orbifold settings, since there one can use modular invariance to compute the one-loop corrections to the gauge kinetic function, and from there the exact superpotential \cite{Font:1990nt,Ferrara:1990ei,Nilles:1990jv}. The racetrack scenario has however the disadvantage of typically leading to vacua with negative cosmological constant. 

Type II orientifold compactifications provide a different scheme. Particularly interesting are Type IIB compactifications with O3/O7-planes. Here we have two different mechanisms that generate a scalar potential, and each of them acts on a different set of light fields, with no kinetic mixing at tree-level between both sets. The first mechanism consists of adding background three-form fluxes, which fix complex structure moduli at a  high scale $M_{\rm flux}$. The second one is the presence of D-brane instantons or perturbative corrections, which stabilize K\"ahler moduli at a lower scale. This structure has led to several proposals to build de Sitter vacua in string theory \cite{Kachru:2003aw,Balasubramanian:2005zx}, whose implementation is currently under debate. One interesting property of this scheme is that three-form fluxes can in principle break supersymmetry at a scale much lower than $M_{\rm flux}$, and that one can compute microscopically the effect of this supersymmetry breaking on the SM sector of the compactification, which is realized on the worldvolume of D-branes \cite{Camara:2003ku,Grana:2003ek,Camara:2004jj}. This implies that, in Type II and F-theory constructions where the MSSM or extensions are localized on a patch of the extra dimensions, one can provide a geometric description of their soft term structure.

Moduli stabilization and supersymmetry breaking scenarios for M-theory on $G_2$ manifolds have been reviewed in \cite{Acharya:2012tw}.

From the macroscopic viewpoint, both the heterotic and Type II models fall into the category of gravity-mediated supersymmetry breaking. This is a general 4d supergravity framework in which the MSSM soft terms are parametrized by F-terms for bulk fields, which have Planck-suppressed couplings to the SM fields \cite{Kaplunovsky:1993rd,Brignole:1993dj}. In the simplest cases, realized by dilaton domination in the heterotic case and by $T$-modulus domination in intersecting D7-branes/F-theory models, all soft terms are specified by only two parameters: the universal gaugino mass and the Higgs $\mu$-term. This simple pattern has the advantage that it can easily avoid unwanted flavor-changing neutral currents, because the flavor patterns in the holomorphic Yukawa sector depend on a set of moduli that decouple from the supersymmetry-breaking sector \cite{Conlon:2007dw}. Imposing experimental constraints on these models (like compatibility with EW symmetry breaking) reduces the two parameters to one, and leads to a scenario where the eventual measurement of a single sparticle mass determines the whole spectrum \cite{Aparicio:2008wh}. In other constructions, like $T$-modulus domination in heterotic and D3-branes at singularities models, the leading-order soft terms vanish, and they only appear through one-loop corrections to the gauge kinetic functions and K\"ahler potential. Such corrections are however quite model dependent, so it is difficult to draw general lessons.

Alternatives to gravity mediated supersymmetry breaking have also been explored in string theory setups, but their full realization remains an open problem. On the one hand, one-loop anomaly mediation requires to implement the idea of {\em sequestering} in the SM sector, which has been explored in Type IIB orientifolds with the SM localized on a warped throat \cite{Kachru:2007xp,Berg:2010ha}. On the other hand, gauge mediation needs to localize the SM in a small region of the extra dimensions and full stabilization of the bulk fields of the compactification at a high scale \cite{Diaconescu:2005pc,Garcia-Etxebarria:2006lri,Argurio:2006ny}. These requirements are in tension with the realizations of the SM  known to date. 

\subsection{Dark Sectors}
\label{sec_darksector}

A general lesson from string theory is that dark sectors are ubiquitous. 
A compactification of string theory generically possesses many more gravitationally coupled degrees of freedom than those in the visible sector attributed to the Standard Model.

This general conclusion can be drawn already from the perturbative heterotic string. For example, if we focus on the heterotic string with gauge algebra $E_8 \times E_8$, the Standard Model can be embedded into one of the two $E_8$ factors, while the second factor acts as a dark or hidden sector.
If the entire Standard Model gauge group is embedded into the first $E_8$ factor, the second $E_8$ factor is uncharged under the Standard Model gauge group and hence does not directly interact at tree-level. 
This sequestering has a particularly appealing geometric interpretation  at strong string coupling, where the theory is better described by M-theory compactified on an interval $S^1/\mathbb Z_2$ with the two $E_8$ being localized on the two end-planes of the interval \cite{Horava:1995qa}.
The interactions between both gauge sectors appear only at the loop level in heterotic string theory, or gravitationally.

The appearance of extra gauge sectors is also a typical property of string compactifications with branes. The total rank of the gauge algebra is constrained by the tadpole  conditions for the branes, which in turn derives from geometric properties of the compactification space. While it is believed that the maximal possible rank that can be obtained from string theory compactifications is finite, it can become extremely high in known examples. For instance, the current record in compactifications of F-theory to six dimensions is ${\rm rk}(G)= 302,896$ and in four dimensions ${\rm rk}(G) > 121,328$ \cite{Candelas:1997pq} (see also \cite{Taylor:2015xtz}).

One might wonder whether the appearance of dark sectors is a moduli dependent statement. If the hidden gauge sector contains massless charged scalar fields, its gauge algebra can in principle be partially higgsed, thereby reducing also the amount of matter, possibly even completely. However, so-called non-Higgsable clusters \cite{Morrison:2012np,Morrison:2014lca}, i.e. gauge sectors with no massless charged scalars that could induce a (partial) Higgsing, seem to be another generic property of the non-perturbative landscape of string vacua. This phenomenon can be studied explicitly in F-theory compactifications: An overwhelming prevalence of non-Higgsable clusters has been observed within a large set of elliptic Calabi-Yau fourfolds \cite{Halverson:2017ffz,Taylor:2017yqr}. 
The glueballs of such confining dark sectors may give rise to interesting dark matter candidates though a typical challenge in this scenario is to prevent the potential oversaturation of the relic abundance. The phenomenological aspects of this scenario have been studied, for instance, in \cite{Halverson:2018olu,Halverson:2020xpg} and references therein.
 This is just one example of dark gauge sectors with potentially interesting phenomenology (see e.g. for a different setup based on strong coupling \cite{Heckman:2011sw}).

The rich sequestered gauge sectors described above arise in addition to the matter sector associated with the higher-dimensional gravitational dynamics itself. Apart from the moduli and axion sector this may potentially even include the Kaluza-Klein towers of light fields such as the graviton. The idea 
of a light Klauza-Klein tower as the origin of dark matter, first suggested in \cite{Dienes:2011ja,Dienes:2011sa}, was recently investigated in \cite{Gonzalo:2022jac} in the context of the Swampland approach to quantum gravity. It rests on the assumption that the smallness of the observed value of the cosmological constant points to our vacuum as arising in the asymptotic regions of a moduli space with one large extra dimension \cite{Montero:2022prj}.
An alternative (and possibly dual) dark matter candidate in such scenarios are primordial black holes, see \cite{Anchordoqui:2022txe} and references therein.

\section{THE STANDARD MODEL AND QUANTUM GRAVITY}

Despite the incredible success of the Standard Model in describing the observed particle physics up to the currently accessible energy scales, there are compelling arguments why it is incomplete. First of all, gravity is not yet incorporated with the other known fundamental forces whose quantum mechanical description underlies the Standard Model. Secondly, numerous astrophysical and cosmological observations have pointed to the existence of dark matter. From a practical standpoint, a complete model of particle physics should account for the interactions with the unknown dark matter that makes up about 27\% of the universe. Thirdly, the mysterious dark energy which dominates the energy density of the current universe defines the vacuum in which we live. A complete theory should explain why our electroweak vacuum has such a minute energy density, and account for this in assessing the electroweak vacuum stability. 
There are other motivations for physics beyond the Standard Model, such as the strong CP problem and the flavor problem, but none has led to structural changes in our thinking of particle physics as the gravitational considerations outlined above.
It would thus seem worthwhile to take seriously the lessons we learned through incorporating realistic particle physics features in string theory.

The electroweak hierarchy problem has been a central focal point 
 in the realm of particle physics, beckoning us to venture beyond the 
 Standard Model.
The hierarchy in question is the unnaturally small ratio of the Higgs mass to the UV cutoff scale (which in the minimal scenario can be identified with the Planck scale).
 Other than aesthetic considerations, there exists a conceptual difficulty associated with the quadratic mass divergences which accompany fundamental scalar fields. 
 These divergences violate a notion of ``naturalness" often attributed to Wilson \cite{Wilson:1970ag}
 which can be roughly stated as a requirement for
the observable properties of a theory to be stable against tiny variations of the fundamental dimensionless parameters.
A notion of ``technical naturalness" in assessing the degree of fine-tuning was further
formulated by 't Hooft \cite{tHooft:1979rat}: ``at any
energy scale $\mu$, a physical parameter or set of parameters
$\alpha_i (\mu)$ is allowed to be very small only if the replacement
of $\alpha_i (\mu)=0$ would increase the symmetry of the system''. 

There are indications that 
gravity 
could offer a departure from this
Wilsonian EFT reasoning.  Heuristic thought experiments with black holes \cite{Banks:2010zn} together with string theoretical arguments based on considerations of the worldsheet conformal field theory \cite{Banks:1988yz} and holography 
\cite{Harlow:2018tng} respectively have substantiated the conjecture that exact global symmetries do not exist in quantum gravity. The lack of global symmetries adds intricacy to the aforementioned limit  $\alpha_i (\mu)
\rightarrow 0$. Another hint that incorporating gravity may lead to a departure from the Wilsonian reasoning is the Beckenstein-Hawking
entropy of a black hole, 
which
connects the classical infrared (IR) solution of gravity with the degeneracy of extremely massive states in the ultraviolet (UV) theory.
This peculiar feature of quantum gravitational theories goes under the umbrella term of UV/IR mixing (a term coined in \cite{Minwalla:1999px}).
 The non-decoupling of UV physics from the IR
manifests in various forms in string theory. For instance, in perturbative string theory, this non-decoupling can be seen as a consequence of modular invariance of the worldsheet partition function. There have been recent attempts,
e.g. \cite{Abel:2021tyt}, to set up the Higgs mass computation in perturbative string
theory 
with this UV/IR duality in mind. 
In what way this UV/IR
mixing concretely addresses the electroweak hierarchy problem
remains to be explored. 

Interestingly, a different notion of naturalness has emerged from studying the vast but finite landscape of
string vacua.
Stringy naturalness
\cite{Douglas:2004zg} 
is a measure of the degree of tuning by the
number of string theory solutions leading to a given value
of an observable. 
Naturalness based on the statistics of string vacua has been used to assess the likelihood of
the scale of supersymmetry breaking
\cite{Susskind:2004uv,Douglas:2004qg,Dine:2004is} 
and in addressing other
vexing hierarchy problems such as the smallness of the cosmological
constant, though explicit string theory realizations of this idea remain to be found.

It was in this backdrop of finding how gravity affects quantum field theory reasonings that the Swampland Program was proposed, named after the term introduced in \cite{Vafa:2005ui}.
The goal of the Swampland Program is to discriminate between EFTs that belong to the Landscape (defined to be the set of theories that can descend from a quantum theory of gravity) and those that do not. See
e.g. \cite{Brennan:2017rbf,Palti:2019pca,vanBeest:2021lhn,Grana:2021zvf,Harlow:2022gzl}
for recent reviews.
A number of quantum gravity constraints (also known as Swampland criteria) have been put forth, mostly by finding patterns in string constructions or through thought experiments involving black holes. The motive behind finding these constraints is to discover the underlying universal principles for consistently coupling quantum field theories to gravity.
 These quantum gravity constraints, if proven, 
can have interesting
phenomenological consequences. For example, milli-charged dark matter
scenarios frequently considered in phenomenological model building are in tension
with the absence of global symmetry in quantum gravity
\cite{Shiu:2013wxa}. The Weak Gravity Conjecture (WGC)
\cite{Arkani-Hamed:2006emk}, which quantifies how much does gravity violate global symmetries, has been used to put phenomenological
constraints on axions \cite{Brown:2015iha, Brown:2015lia,
Montero:2015ofa,Heidenreich:2015wga,Hebecker:2018ofv} and dark
photons \cite{Reece:2018zvv}. 
There have been attempts to leverage the WGC in the presence of scalars to set an upper bound on the weak scale  \cite{Lust:2017wrl,Craig:2019fdy}.
A stronger versions of the WGC which presumes the absence of stable non-supersymmetric AdS vacua \cite{Ooguri:2016pdq} has been
used to link the observed value of the cosmological constant with the
neutrino masses (which for fixed Yukawa couplings, set the weak scale)
\cite{Ooguri:2016pdq,Ibanez:2017kvh,Hamada:2017yji}. 
It should be noted that such arguments
hinge on being able to demonstrate that the AdS vacua in question are stable. 
Without knowledge of the full UV spectrum, we cannot exclude the possibility of lower energy solutions. In fact, the Standard Model itself already contains a source of instability. Charged matter like the electron generates a potential for the Wilson line moduli (originated from the $U(1)_{EM}$
of the Standard Model upon compactification). It was pointed out in \cite{Hamada:2017yji} that the potential after taking into account the Wilson line moduli has a runaway behavior.
Gravitational positivity bounds \cite{Hamada:2018dde,Bellazzini:2019xts,Alberte:2020jsk, Tokuda:2020mlf} obtained through studies of scattering amplitudes have recently been used to constrain new physics models involving massive gauge bosons \cite{Aoki:2023khq}.
Since the focus of this review is to extract lessons we learned from top-down constructions rather than bottom-up Swampland considerations, we refer the readers to some recent reviews \cite{Cvetic:2022fnv, Draper:2022pvk, deRham:2022hpx} for an in-depth discussion of other phenomenological consequences of the Swampland idea.

As we have seen in the previous sections, the gauge and matter content of top-down string constructions are tightly constrained. An interesting question is whether the constrained spectra are artifacts of looking under the lamppost, or there are general principles that forbid certain particles or interactions to arise in a consistent quantum theory of gravity.
Recent works have gathered support for the latter.
The
Completeness Hypothesis \cite{Polchinski:2003bq} asserts that physical
states with all possible gauge charges consistent with Dirac
quantization are present in a consistent theory of quantum gravity. It has since been found that the Completeness Hypothesis follows from the absence of generalized non-invertible global symmetries \cite{Gaiotto:2014kfa,Harlow:2018tng,Rudelius:2020orz,Heidenreich:2021xpr}.
A corollary of this hypothesis when applied to states charged under 1-form gauge fields is the existence of magnetic monopoles. 
When higher-form symmetries come into play, the
physical states in question would include extended objects, such as strings and branes.
A novel approach using brane probes to test the consistency of EFTs coupled to gravity was initiated in
 \cite{Kim:2019vuc} where it was shown that unitarity on the brane probes can
rule out infinite families of anomaly free
gravitational theories. 
Further studies have found precise match of
the allowed spectra with string constructions, sometimes involving
the global structure of the spacetime gauge group
\cite{Cvetic:2020kuw,Montero:2020icj,Font:2020rsk,Cvetic:2021sjm,Font:2021uyw,Cvetic:2021sxm,Bedroya:2021fbu,Cvetic:2022uuu}.
Taken together, these findings suggest a notion of string universality
\cite{Adams:2010zy} (or the string lamppost principle
\cite{Montero:2020icj}) that all consistent 
theories of
quantum gravity are realized in string theory. 
It is conceivable that exploring this avenue of research to gravitational theories in lower
dimensions and with reduced supersymmetry
may shed light on the constrained spectra 
observed in four-dimensional string vacua that bear significance in phenomenological contexts.
Before then,
 six-dimensional
supergravity theories and their string/F-theory realizations are
an intermediate testbed for substantiating this approach.
This is because this class of string
constructions is fairly well understood,
and known
quantum consistency conditions already tightly bound the set of possible
low-energy theories.
See \cite{Morrison:2021wuv,Raghuram:2020vxm} for investigations of the Completeness Hypothesis and related quantum gravity constraints in this controlled setting.  

We hope we have conveyed  that further understanding of
quantum gravity constraints, in tandem with continued advances in string compactifications,
has the potential to reveal the inherent structure found in the particle spectra prevalent in string theory.

\section{WHAT HAVE WE LEARNED?}
We would like to conclude this brief review by returning to the question posed in its title: What have we learned from Standard Model constructions in string theory?
\begin{enumerate}
\item {\it The SM can be embedded into string theory:}
The main ingredients of the SM - the gauge group, chiral matter and Yukawa interactions - follow relatively naturally from general principles of string theory.
Reproducing {\it exactly} the SM including, for instance, its vector-like matter content or the precise flavor structure, is more involved, but each property per se has been obtained or is within reach. The challenge is to find vacua that combine all these features at the same time, with all moduli stabilized and incorporating a realistic cosmology.
\item {\it Why the world is described by the SM remains mysterious:} Why the SM gauge group, matter content (e.g. three chiral families) or couplings describe our world is to date not clear from a string theoretic point of view, beyond anthropic arguments not specific to string theory, and may not even have a good explanation. This is related to the vacuum selection problem which is yet to be deciphered in the future.
\item {\it String theory can inform model building and sheds new light on naturalness:}
String theory offers solutions to model building challenges beyond four-dimensional quantum field theory: 
It can incorporate higher-dimensional mechanisms to generate hierarchies such as warped extra dimensions in a computable and UV complete framework. It provides mechanisms to generate hierarchical Yukawa couplings, topological mechanisms to break GUT symmetries and offers solutions to associated model building challenges.
A goal of string phenomenology for the future is to identify more such general stringy mechanisms. String theory also naturally includes rich dark sectors and oftentimes comes with relatively light extra scalar fields and axions, at least in corners of the landscape under current computations control. 
\item {\it Quantum gravity is essential to understand particle physics:} Gravitational physics plays a more important role for particle model building than naively expected, and string theory, unlike purely field theoretic approaches to phenomenology, has all of the ingredients to help us understand this profound connection. For instance, all aspects of Abelian or discrete gauge symmetries and their associated selection rules are sensitive to the gravitational sector and cannot be studied in isolation. More generally this is a key lesson from the Swampland Program. Harvesting the constraints from quantum gravity for particle physics via string theory as a computationally well defined framework furnishes a promising research direction for modern string phenomenology.
\end{enumerate}


\section*{ACKNOWLEDGMENTS}
We thank the editors of Annual Review of Nuclear and Particle Science for the invitation to write this short review, and Michael Peskin for useful comments.
F.\,M.\ is supported through the grants CEX2020-001007-S and PID2021-123017NB-I00, funded by MCIN/AEI/10.13039/501100011033 and by ERDF A way of making Europe.
G.\,S.\ is supported in part by the DOE grant DE-SC0017647.
 T.\,W.\ is supported in part by Deutsche Forschungsgemeinschaft under Germany's Excellence Strategy EXC 2121  Quantum Universe 390833306 and by Deutsche Forschungsgemeinschaft through a German-Israeli Project Cooperation (DIP) grant ``Holography and the Swampland”.

\printbibliography

\end{document}